\documentclass{ws-procs975x65}

\begin{document}

\title{Finlser geometric extension of Einstein gravity}

\author{C. Pfeifer$^*$ and M. Wohlfarth}

\address{II Institute for theoretical physics, University of Hamburg,\\
Hamburg, Germany\\
$^*$E-mail: christian.pfeifer@desy.de}

\begin{abstract}
We present our Finsler spacetime formalism which extends the standard formulation of Finsler geometry to be applicable in physics. Finsler spacetimes are viable non-metric geometric backgrounds for physics; they guarantee well defined causality,
the propagation of light on a non-trivial null structure, a clear notion of physical observers and the existence of physical field theories determining the geometry of space-time dynamically in terms of an extended gravitational field equation. 
Here we give the precise definition of Finsler spacetimes and the notion of well-defined observers, their measurements, the transformations between them and how to formulate action based field theories.
\end{abstract}


\section{Finsler spacetimes}
The standard textbook formulation of Finsler geometry \cite{Chern} is a well known extension of Riemannian metric geometry based on a general length measure $S$ for curves $\gamma$
\begin{equation}
 L[\gamma]=\int F(\gamma,\dot\gamma),
\end{equation}
rather then on a positive definite metric $g$ which induces the length measure via $F=\sqrt{g_{ab}(\gamma)\dot\gamma^a\dot\gamma^b}$. It has the shortcoming that it can not be used as a generalization of Lorentzian metric spacetime geometry,
since the geometric objects like connections and curvature are not well-defined in case $F$ has a non-trivial null structure. Our definition of Finsler spacetimes solves this problem and sets the stage for the application of non-metric Finslerian 
spacetime geometry in physics by introducing a smooth n-homogeneous fundamental geometry function $L(x,y)$ on the tangent bundle. For the tangent bundle TM of the spacetime manifold M we use manifold induced 
coordinates $(x,y)=Z\in TM, Z=y^a\partial_a{}_{|x}$ and $\{\partial_a=\frac{\partial}{\partial x^a},\bar \partial_a=\frac{\partial}{\partial y^a}\}$ as basis of $T_{(x,y)}TM$, see our articles \refcite{Pfeifer:2011xi,Pfeifer:2011tk} for all details.

\textit{A Finsler spacetime $(M,L,F)$ is a smooth manifold~$M$ and a smooth function $L$ on the tangent bundle, homogeneous of degree $n$ with respect to $y$ such that:
\begin{enumerate}[(i)]
\item $L$ is reversible $|L(x,-y)|=|L(x,y)|$,
\item $g^L_{ab}=\frac{1}{2}\bar\partial_a\bar\partial_b L$ is non-degenerate on $TM\setminus A$; $A\subset TM$ of measure zero,
\item $\forall x\in M$ there exists a non-empty closed connected set $S_x\in T_xM$ where: $|L(x,y)|=1$ and $sign( g^L{}_{ab})=(\epsilon,-\epsilon,-\epsilon,-\epsilon)$ with $\epsilon=\frac{|L(x,y)|}{L(x,y)}$
\end{enumerate}
The Finsler function is $F(x,y) = |L(x,y)|^{1/n}$, the Finsler metric $g^F_{ab}=\frac{1}{2}\bar\partial_a \bar\partial_b F^2$.}

\begin{figure}[h!]
  \begin{center}
  \includegraphics[width=0.4\textwidth]{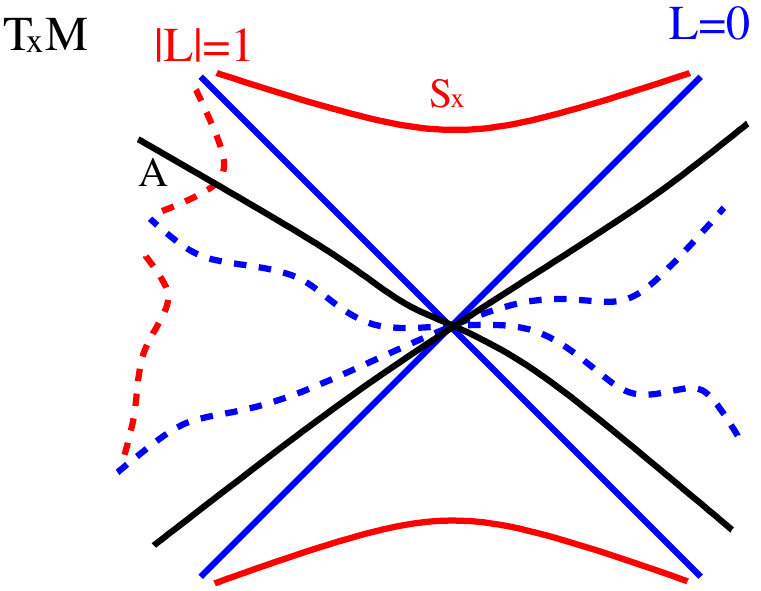}
  \end{center}
  \caption{\label{fig:fst}Causal structure of Finsler spacetime.}
\end{figure}

Our definition of Finsler spacetimes guarantees a causal structure in each tangent space: $S_x$ is the shell of unit timelike vectors which defines a cone of timelike directions with null boundary, as displayed in figure \ref{fig:fst}.

The geometry of Finsler spacetimes is solely derived from derivatives of $L$ in terms of the unique Cartan non-linear connection coefficients: $N^a{}_b=\frac{1}{4}\bar\partial_b(g^{Laq}(y^m\partial_m\bar\partial_qL-\partial_qL))$. 
The connection between our definition of Finsler spacetimes and standard Finsler geometry is given by the following theorem: \textit{Wherever L and F are both differentiable they encode the same geometry, i.e. $N[L]=N[F^2]$}. 
The fact that Finsler spacetimes are built from a smooth function $L$ enables us to consider for example geometries based on $L=G_{a_1...a_n}y^{a_1}...y^{a_n}$ with non-trivial null-structure. In standard Finsler geometry
this would not be possible since the geometric objects based on derivatives of $F=(G_{a_1...a_n}y^{a_1}...y^{a_n})^{1/n}$ would not be well defined on the null-structure of spacetime.

\section{Observers}
\begin{figure}[h]
  \begin{center}
  \includegraphics[width=0.4\textwidth]{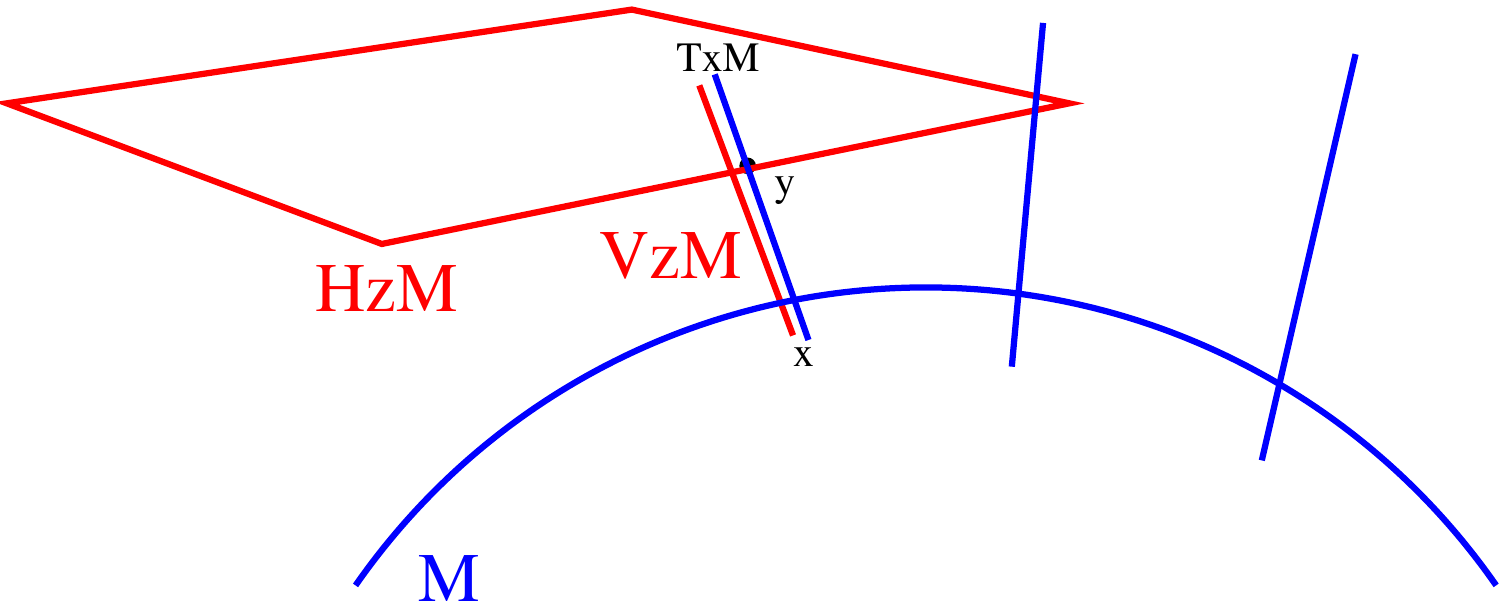}
  \includegraphics[width=0.4\textwidth]{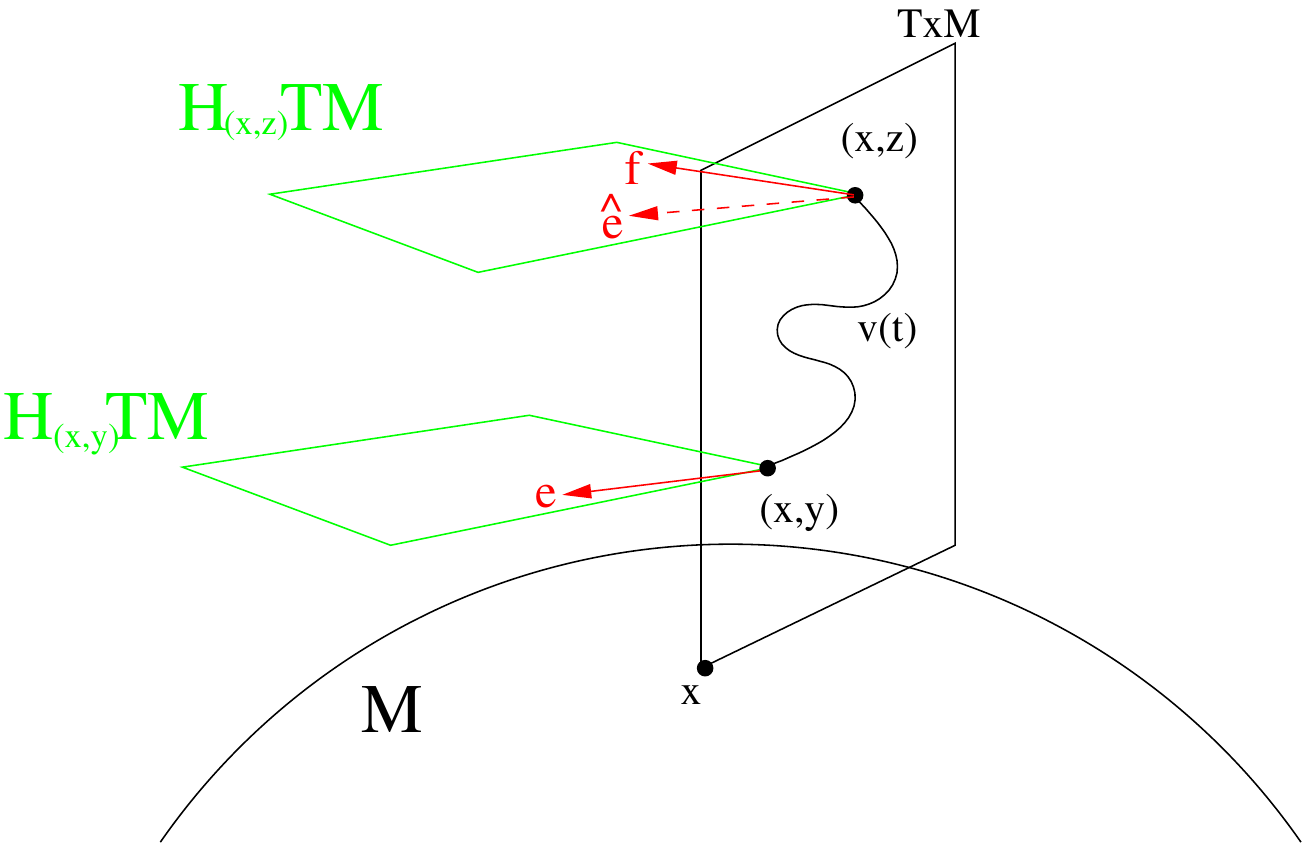}
  \end{center}
  \caption{\label{fig:hv}Left: Horizontal and vertical tangent space to the tangent bundle.}
  \caption{\label{fig:obstraf}Right: Transformations between observers.}
\end{figure}

The connection coefficients $N$ split TTM and T*TM into horizontal and vertical space by $\{\delta_a=\partial_a-N^b{}_{a}\bar\partial_b,\bar\partial_a\}$ and $\{dx^a,\delta y^a=dy^a+N^a{}_{v}dx^b\}$, as displayed in figure \ref{fig:hv}. 
The horizontal (co-)tangent space is identified with the (co-)tangent space along the manifold directions. Timelike observers move on worldlines $x(\tau)\in M$ with trajectory $(x, \dot x)\in TM$ and $\dot x$ in the cone of timelike vectors. A 
horizontal orthonormal frame defines their time and space directions along the manifold $\{e_a\}=\{e_0=\dot x^a\delta_a,e_\alpha\}; g^F_{(x,\dot x)}(e_\mu,e_\nu)=-\eta_{\mu\nu}$. Measurable quantities are components of horizontal tensors evaluated 
in this frame at the observers $TM$ position. 

Frames of different observers $e$ and $f$ are located in different tangent spaces to $TM$, even if they are at the same position of spacetime, see fig. \ref{fig:obstraf}. Therefore the transformation from one observer frame $e$ to the other $f$ 
involves first parallel transport of frame $e$, to the tangent bundle position of the other observer frame $f$ along the vertical geodesic $v(t)$. There the transported frame $\hat e$ and $f$ are living in the same tangent space to TM and can 
therefore be mapped onto each other by a linear transformation, which turns out to be a Lorentz transformation. The combination of parallel transport and Lorentz transformation form a groupoid. All technical details and the proofs of the statements
can be found in Ref. \refcite{Pfeifer:2011xi}. 

\section{Field theories and gravitational dynamics}
The geometry of Finsler spacetimes is built from tensors on $TM$; hence physical fields coupling to this geometry will be of the same kind. Lagrange densities on $TM$ require the canonical Sasaki-type $TM$-metric 
$G=-g^F{}_{ab}(dx^adx^b+F^{-2}\delta y^a\delta y^b)$, which allows us to couple field theories to Finsler spacetime geometry as follows: Pick an action for a $p$-form $\phi(x)$ on $(M,g): S[\phi, g]=\int_M\sqrt{g}\mathcal{L}(g,\phi,d\phi)$, 
use the Lagrangian for a zero homogeneous $p$-form field $\Phi(x,y)$ on $(TM,G)$, introduce Lagrange multipliers to restrict the $p$-form field to be horizontal, integrate over the unit tangent bundle $\Sigma=\{(x,y)\in TM | F(x,y)=1\}$ to obtain 
the $p$-form field action $S_m[\Phi, L, \lambda]=\int_\Sigma(\sqrt{g^Fh^F} \mathcal{L}(G,\Phi,d\Phi)+\lambda(1-P^H)\Phi)_{|\Sigma}$. Our coupling principle ensures that in case the Finsler spacetime is metric, field theories and gravitational dynamics equal those of general relativity.

The geodesic deviation on Finsler spacetimes gives rise to a tensor causing relative gravitational acceleration $\nabla_{\dot x}\nabla_{\dot x}V^a=R^a{}_{bc}(x,\dot x)\dot x^bV^c$. This non-linear curvature is built from the non-linear connection 
coefficients leads to the curvature scalar $\mathcal{R^F}=R^a{}_{ab}y^b$ which we choose as Lagrangian for our Finsler gravity action
$S[L,\Phi]=\int_\Sigma(\sqrt{g^Fh^F}\mathcal{R^F})_{|\Sigma}+S_m[L,\Phi]$. Variation with respect to the $L$ yields the Finsler gravity field equation 
\begin{equation}
g^{Fab}\bar\partial_a\bar\partial_bR^F-\frac{6}{F^2}R^F+2g^F_{ab}\big(\nabla_aS_b+S_aS_b+\bar\partial_a(y^q\nabla_q S_b)\big)=-\kappa T_{|\Sigma}\,.
\end{equation}
In case the function L is the metric length measure the Finsler gravity equation is equivalent to the 
Einstein equations. Details and a first order perturbative solution of the Finsler gravity equation can be found in Ref. \refcite{Pfeifer:2011xi}.

\bibliographystyle{ws-procs975x65}
\bibliography{main}

\end{document}